\documentclass[review,11pt]{elsarticle}

\usepackage{amssymb}
\usepackage{enumitem}
\usepackage{lineno}
\usepackage{multirow}
\usepackage{rotating}
\usepackage{epstopdf}
\usepackage{graphicx}
\usepackage{subfigure}

\usepackage{times}
\usepackage{booktabs} 

\usepackage{hyperref}
\hypersetup{
	pdftitle={Degree Distribution of Delaunay Triangulations},
	pdfauthor={Gang Mei}
}

\journal{arXiv.org}

\begin{document}
\begin{frontmatter}
		
	\title{Degree Distribution of Delaunay Triangulations}


\author[label1]{Gang Mei\corref{cor1}}
\ead{gang.mei@cugb.edu.cn}
\author[label1]{Nengxiong Xu}
\author[label2]{Salvatore Cuomo}

\cortext[cor1]{Corresponding author}
\address[label1]{School of Engineering and Technology, China University of Geosciences (Beijing), 100083, Beijing, China}
\address[label2]{Department of Mathematics and Applications ``R. Caccioppoli'', University of Naples Federico II, Naples, Italy}

\begin{abstract}
Delaunay triangulation can be considered as a type of complex networks. For 
complex networks, the degree distribution is one of the most important 
inherent characteristics. In this paper, we first consider the two- and 
three-dimensional Delaunay Triangulations (DTs) as a type of complex 
networks, and term it as DT networks. Then we statistically investigate the 
degree distribution of DT networks. We find that the degree distribution of 
DT networks well follows the Gaussian distribution in most cases, which 
differs from the Poisson distribution and the Power-Law distribution for the 
well-known Small-World networks and Scale-Free networks.
\end{abstract}


\end{frontmatter}



\section{Introduction}
\label{sec:Intro}
Quite recently, an exciting application of the Artificial Intelligence (AI) \cite{1} is that: scientists have used Deep Learning algorithms \cite{2} to recreate the complex neural codes that the place cells and 
grid cells in the brain use to navigate through space \cite{3}. The grid 
cells discovered by May-Britt Moser and Edvard Moser \cite{4} are 
capable of providing a strikingly periodic representation of self-location 
and navigation in the space \cite{5}, which spatially distributes as the 
hexagonal lattices. An interesting phenomenon is that: if considering each 
grid cell as a distinct point and then creating a triangulation for those 
scattered points, then a nice triangulation can be generated since each 
point has six neighboring points and triangles. In a triangulation (more 
precisely, a triangular mesh in this case), the quality of each triangle 
could be very high when each point has six neighboring points and triangles 
since the triangles could be close to the \textit{Regular Triangles}. In other words, when each point 
in a triangulation has the degree of six, a very high-quality triangulation 
can be generated. 

When using a triangulation to represent the spatial connection of the grid 
cells, a network (or so-called complex network) could be achieved. In the 
network of grid cells, the degree of each vertex is in general six. That is, 
the grid cell network represented with a triangulation is a 6-regular 
network with uniform degree distribution. However, for other complex 
networks such as the biological networks or the social networks, most of 
them are not regular networks but the so-called Small-World networks 
\cite{6} or the Scale-Free networks \cite{7}. The degree 
distributions of the Small-World networks and the Scale-Free networks are 
not uniform, but generally follow the Poisson distribution and the Power-Law 
distribution, respectively. 

Inspired by representing the grid cells with triangulations, we consider a 
specific category of the triangulations, i.e., the Delaunay Triangulation 
(DT) \cite{8}, as a type of complex networks. We term this network as the \textit{DT network}. More 
specifically, we consider the two-dimensional DT (i.e., the triangular mesh) 
and the three-dimensional DT (i.e., the tetrahedral mesh) as a specific type 
of complex networks; and we ignore those higher-dimensional DTs. 
Furthermore, we are quite interested in statistically investigating the 
degree distribution of the DT networks. To the best of the authors' 
knowledge, there is no previous work specifically focusing on this problem.

Our work in this paper can be summarized as follows. First, we consider the 
two- and three-dimensional DTs as a specific type of complex networks that 
is termed as the DT network. Second, we statistically investigate the degree 
distribution of the DT network, and find that the degree distribution 
follows a \textit{Gaussian} function, while in contrast the degree distributions of two 
well-known networks, i.e., the Small-World networks \cite{6} and the 
Scale-Free networks \cite{7} follow a \textit{Poisson} function and a \textit{Power-Law} function, 
respectively. 

\section{Method}
\label{sec:Method}
\subsection{Considering Delaunay Triangulations as A Type of Complex 
	Networks}

In mathematics, for a set P of points in the $d$-dimensional Euclidean space, a 
Delaunay Triangulation \cite{8} is a triangulation DT(P) such that no point in 
$P$ is inside the circum-hypersphere of any d-simplex in DT(P). It is known \cite{8}
that there exists a unique Delaunay triangulation for P if P is a set 
of points in general position. For a set of two-dimensional points P, the 
DT(P) is a triangulation such that no point in P is inside the circumcircle 
of any triangle in DT(P). For a set of three-dimensional points P, the DT(P) 
is a triangulation such that no point in P is inside the circumscribing 
sphere of any tetrahedron in DT(P).

\begin{figure}[ht]
	\centering
	\includegraphics[width=\textwidth]{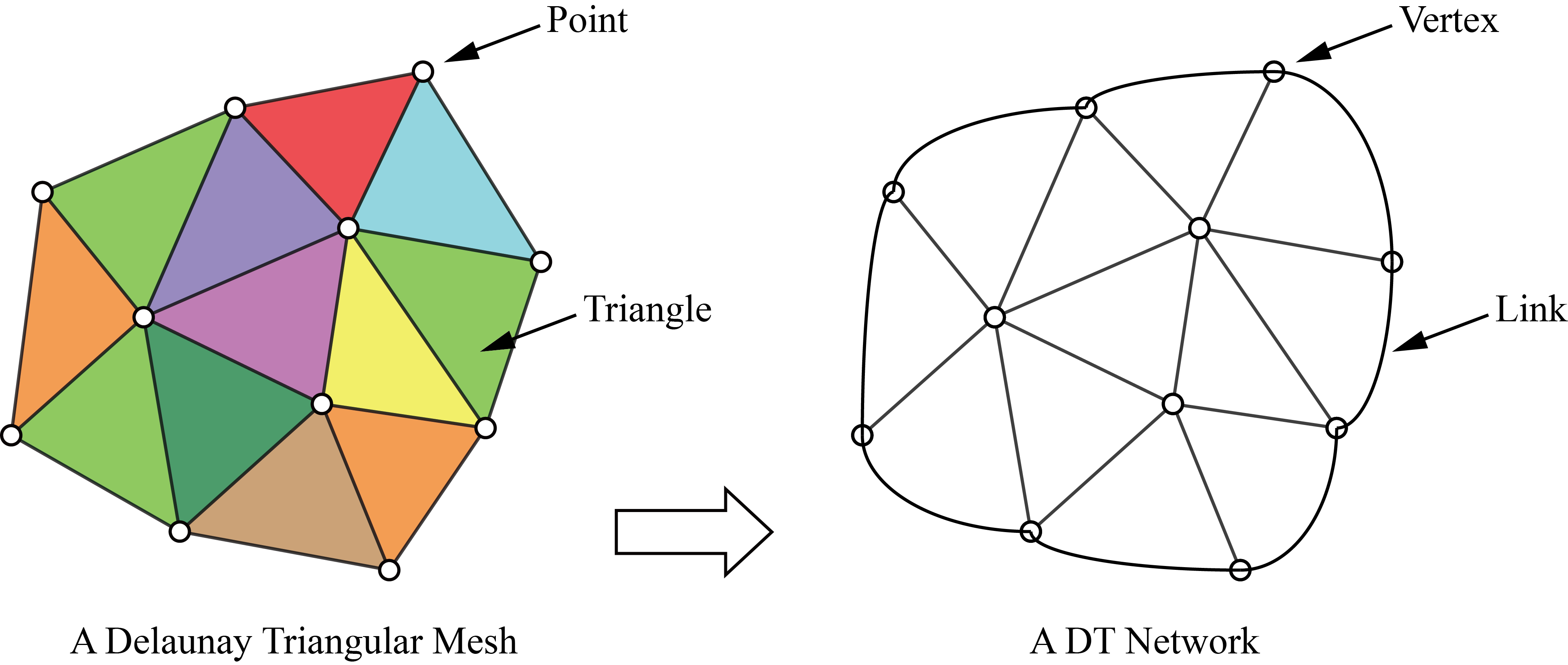}
	\caption{Consideration of a DT as a complex network}
	\label{fig:1}       
\end{figure}

In the context of network theory, a complex network is a graph (network) 
with non-trivial topological features. In this paper, we consider the two- 
and three-dimensional DTs \cite{8}, i.e., the Delaunay triangular meshes and the 
Delaunay tetrahedral meshes, as a specific type of complex networks. More 
precisely, we consider that: each vertex in a DT represents a cell, a 
person, or even a device in a complex network, while each edge in a DT 
represents the link or connection between vertices; see Figure \ref{fig:1}.

\subsection{Investigating the Degree Distribution of DT Networks}

In this paper, we are interested in statistically investigating the degree 
distribution of two- and three-dimensional DT networks. The DT networks are 
in fact the Delaunay triangular meshes in two-dimensions and the Delaunay 
tetrahedral meshes in three-dimensions. To statistically examine the degree 
distribution of DT networks, we first create five groups of triangular 
meshes and tetrahedral meshes based on sets of scattered points. Each set of 
scattered points are randomly generated, and then are used to create the 
triangular or tetrahedral meshes using the two famous Delaunay mesh 
generators, Triangle \cite{9} or TetGen \cite{10}. 

For each DT network, we first obtain the vertex degree (i.e., the number of 
neighboring vertices) for each vertex, and then count the degrees of all 
vertices to obtain the discrete frequency of vertex degree. The frequency of 
vertex degree is represented in percentage. Finally, the frequency of vertex 
degree is fitted to a Gaussian function (Eq.(\ref{eq1})):
\begin{equation}
\label{eq1}
y=ae^{-\frac{(x-b)^2}{2c^2}}+y_0 ,
\end{equation}
where, the parameters $a$, $b$, $c,$ and $y_0 $ are arbitrary real numbers.. The graph 
of a Gaussian is a characteristic "bell curve" shape. The parameter $a$ is the 
height of the curve's peak; $b$ (the mean) is the position of the center of the 
peak; $c$ (the standard deviation) controls the width of the "bell"; and $y_0 $ 
is the offset. 

After fitted the frequency of vertex degrees to the Gaussian model, we use 
the R-Square (R$^{2})$ and Adjusted R-Square (Adj. R$^{2})$ to evaluate the 
Goodness of Fit. 

\section{Results and Discussion}

\subsection{Degree Distribution of Two-dimensional DT Networks}

\begin{figure}[!h]
	\centering
	\subfigure[A DT network with 100 vertices]{
		\label{fig:2:a}       
		\includegraphics[width=0.46\textwidth]{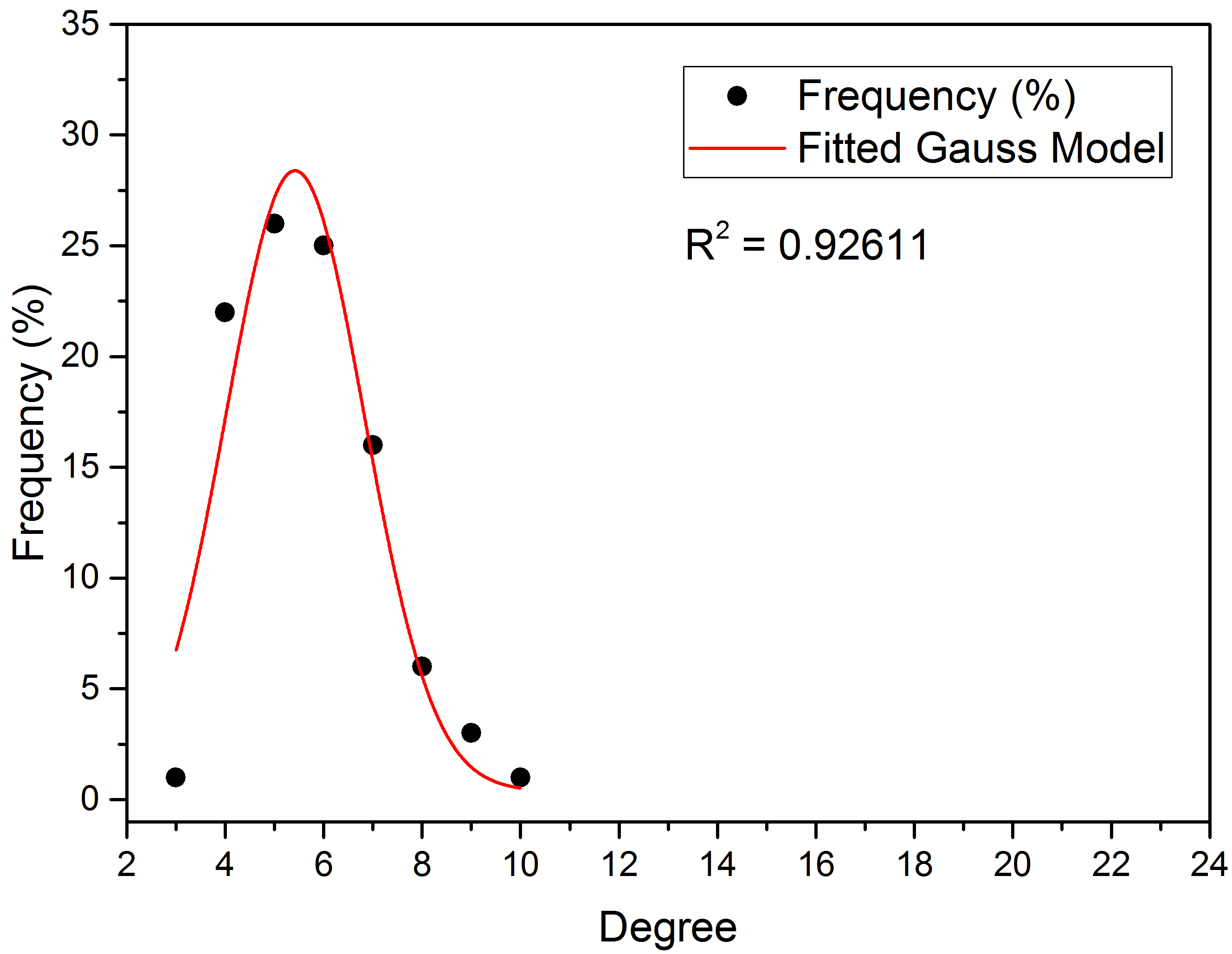}
	}
	\hspace{1em}
	\subfigure[A DT network with 1000 vertices]{
		\label{fig:2:b}       
		\includegraphics[width=0.46\textwidth]{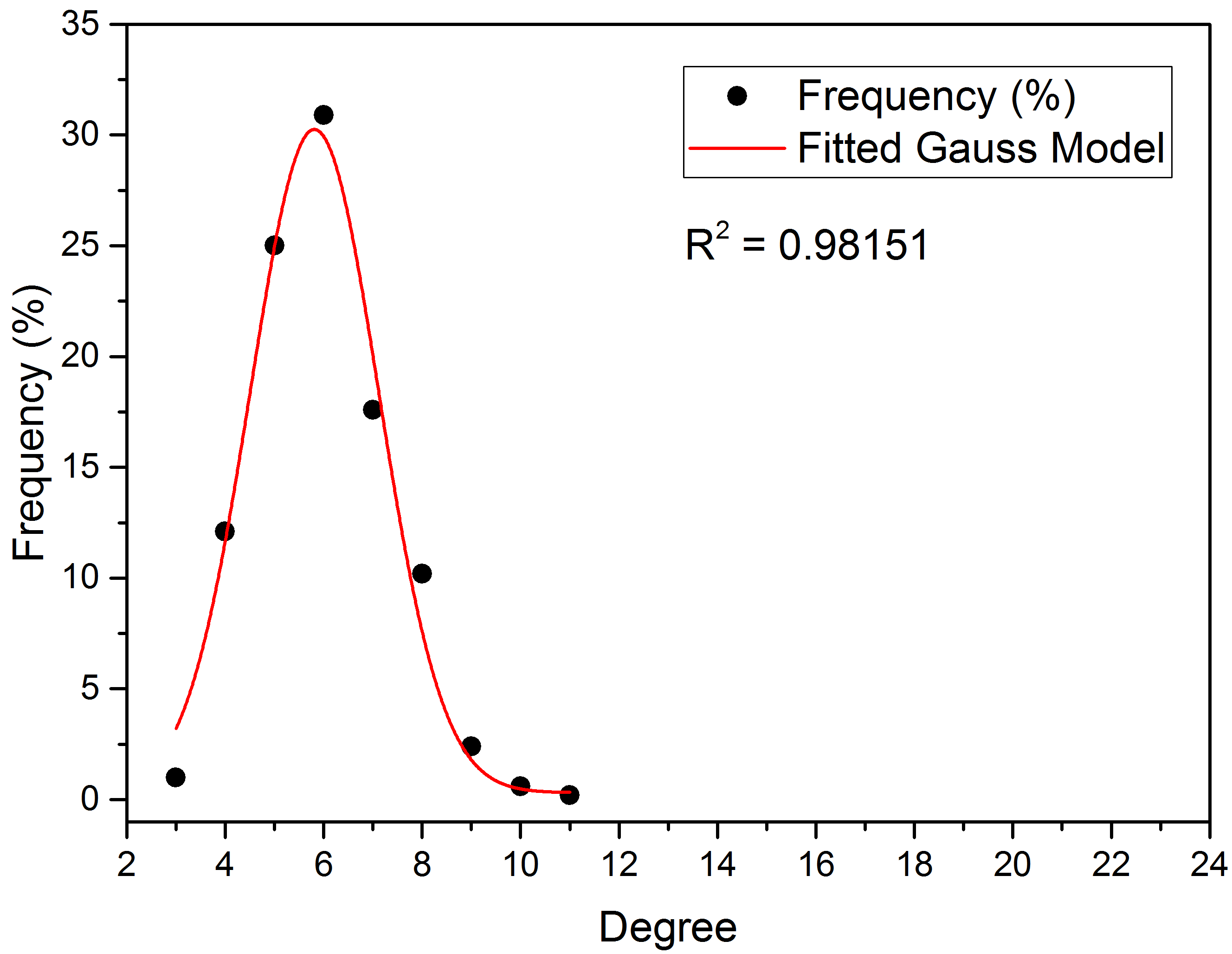}
	}
		\subfigure[A DT network with 10,000 vertices]{
			\label{fig:2:c}       
			\includegraphics[width=0.46\textwidth]{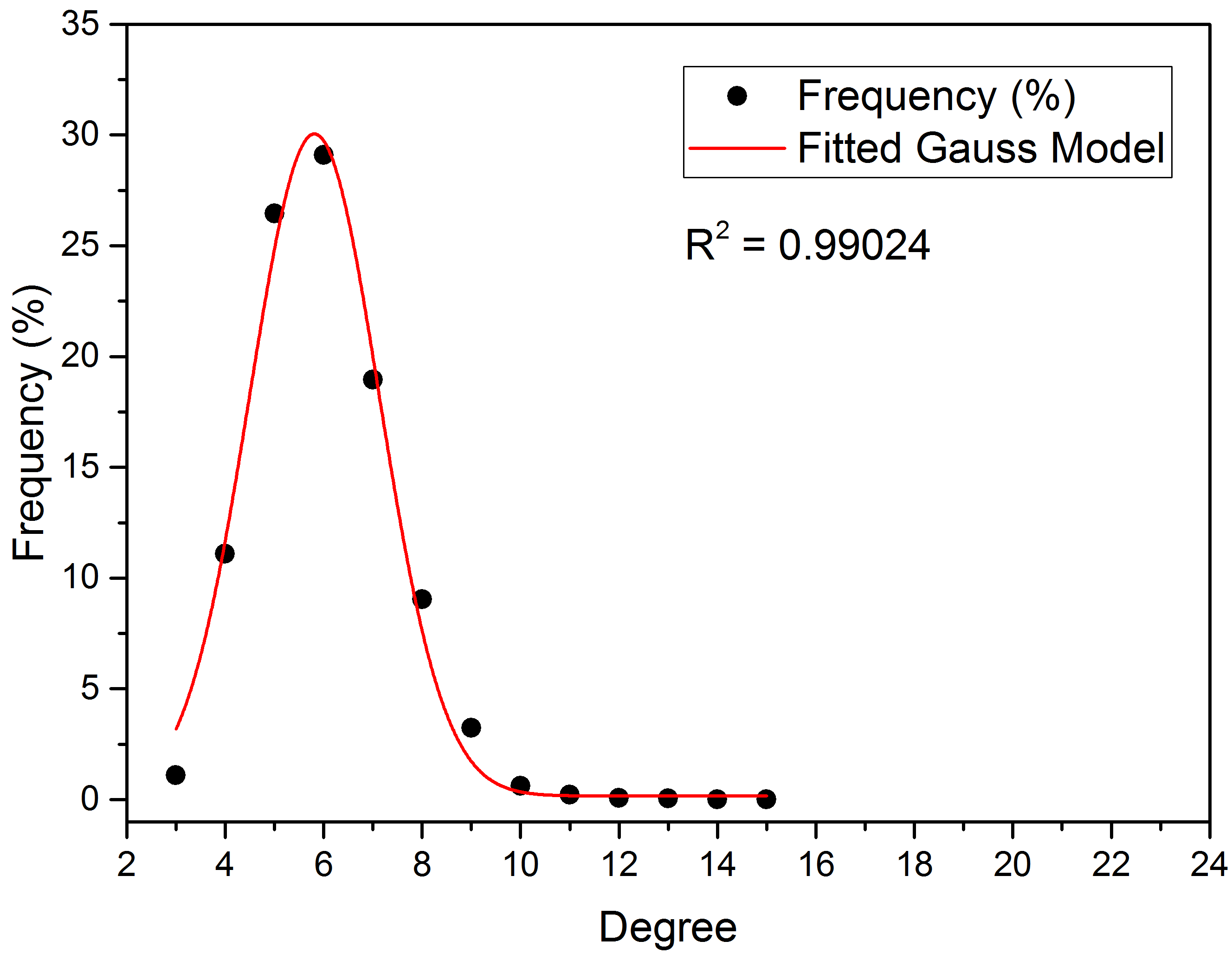}
		}
		\hspace{1em}
		\subfigure[A DT network with 100,000 vertices]{
			\label{fig:2:d}       
			\includegraphics[width=0.46\textwidth]{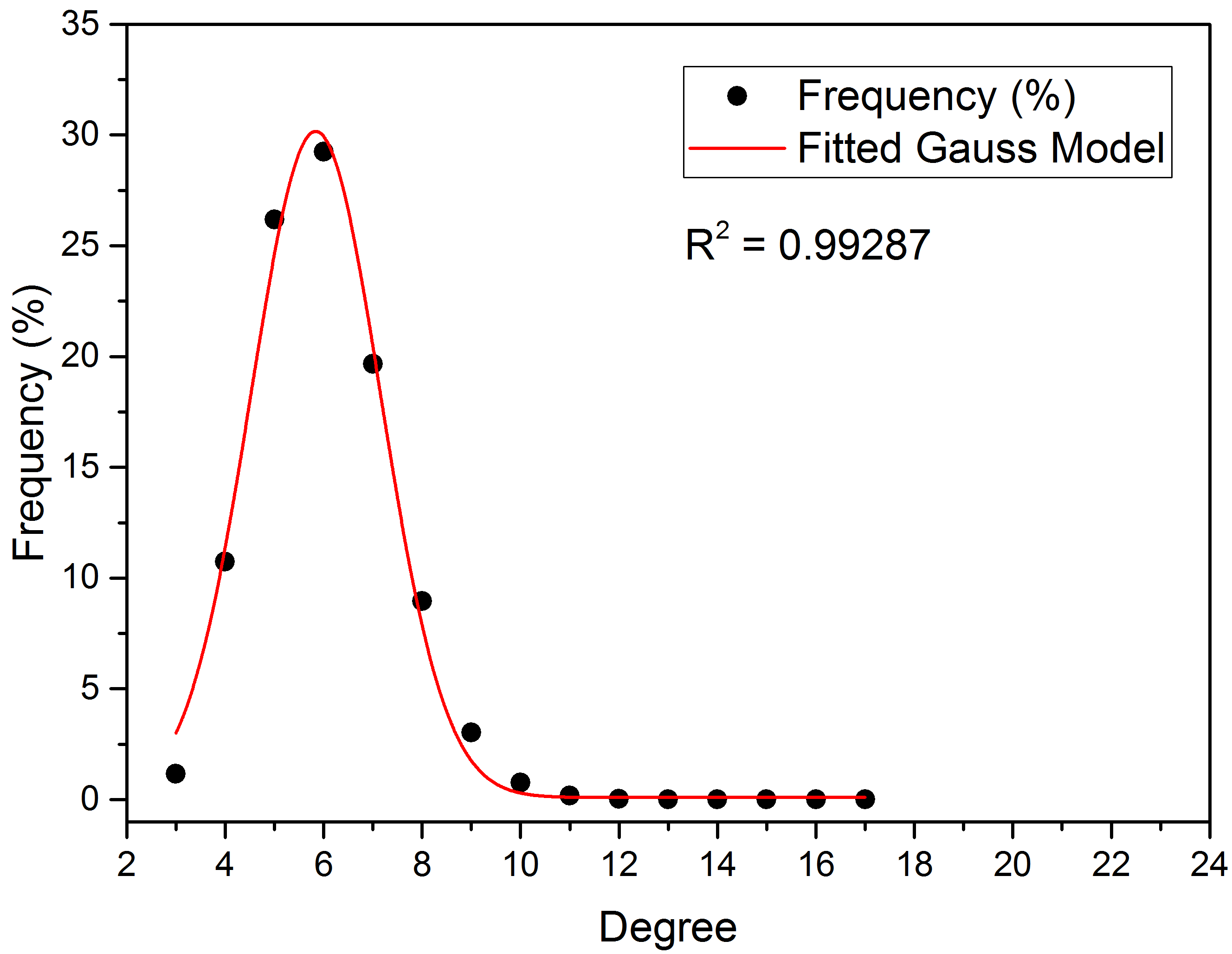}
		}
			\subfigure[A DT network with 1,000,000 vertices]{
				\label{fig:2:e}       
				\includegraphics[width=0.46\textwidth]{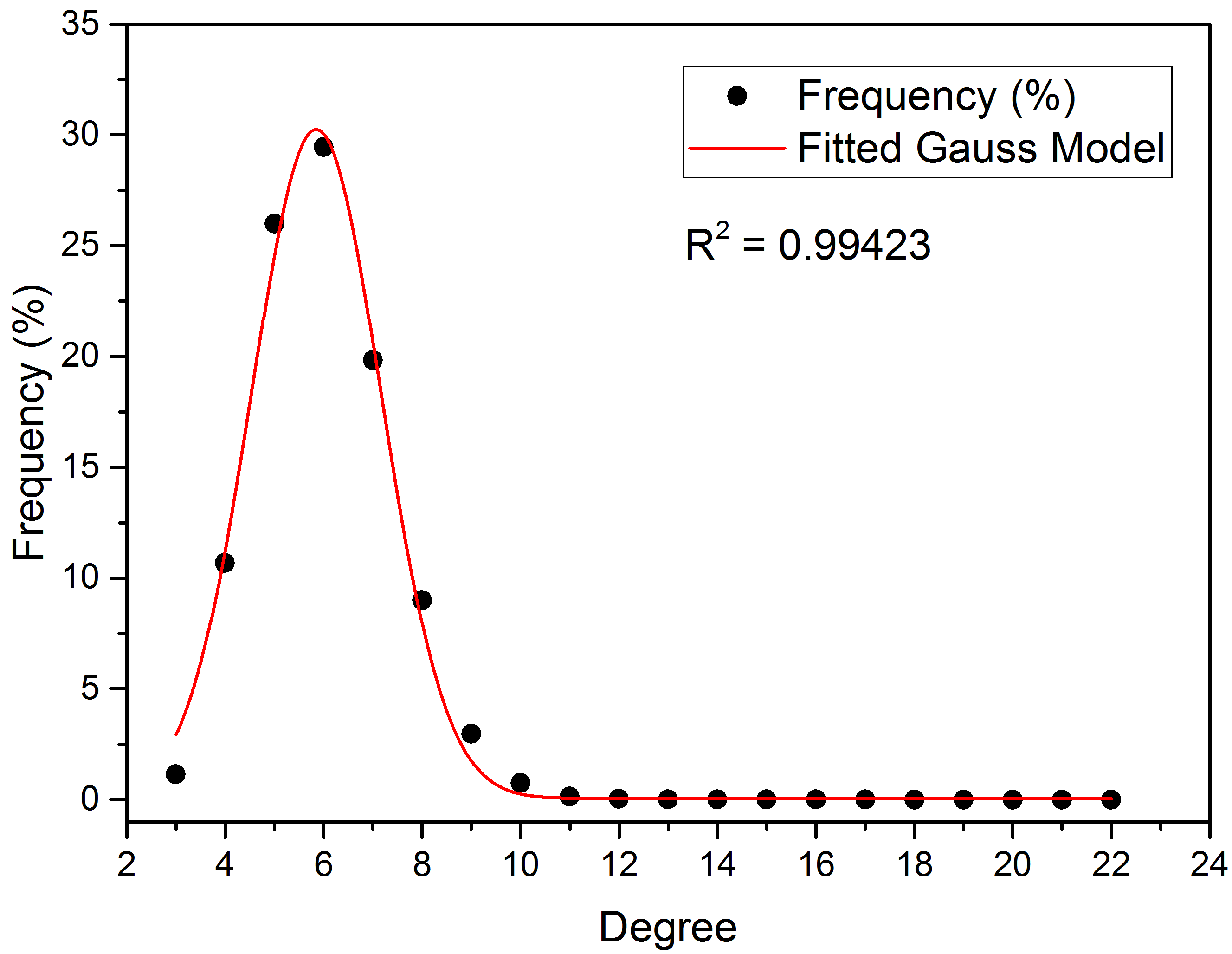}
			}
	\caption{Fitted Gaussian models for five different sizes of 
		two-dimensional DT networks}
	\label{fig:2}       
\end{figure}

We first determine five different sizes of sets of scattered points (i.e., 
100, 1000, 10000, 100000, and 1000000). For each size, we repeatedly create 
five two-dimensional Delaunay triangulations. And there are totally 25 
Delaunay triangular meshes. For all of the 25 meshes, we calculate the 
vertex degrees, count the frequency of vertex degree, and fit the frequency 
of vertex degree to the Gaussian Model (Eq. (\ref{eq1})). We employ the R-Square 
(R$^{2})$ and Adjusted R-Square (Adj. R$^{2})$ to measure the Goodness of 
fit. Details of the fitted Gaussian Models are listed in Table \ref{tab1}. For each 
size, we also plot one of the fitted Gaussian Models (see Figure \ref{fig:2}).

The results presented in Table \ref{tab1} and Figure \ref{fig:2} indicate that: 

\begin{enumerate}[label=(\arabic*), leftmargin=*]
	\item the degree distribution of two-dimensional DT network well follows the Gaussian Distribution in most cases;
	\item with the increase of the number of vertices in DT networks, the fitting of the frequency of vertex degree to a Gaussian model becomes better.
\end{enumerate}

\begin{figure}[!h]
	\centering
	\subfigure[A DT network with 100 vertices]{
		\label{fig:3:a}       
		\includegraphics[width=0.46\textwidth]{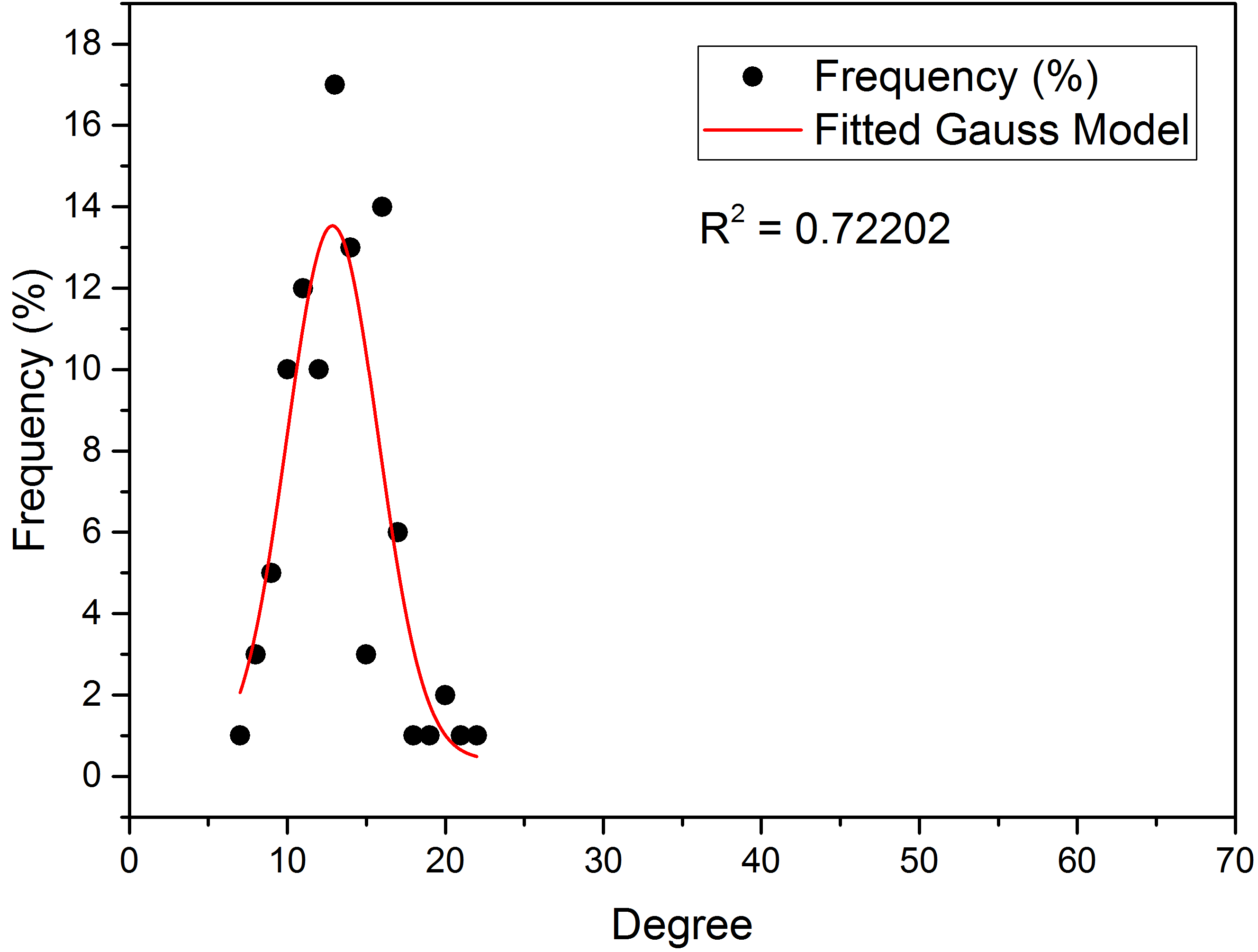}
	}
	\hspace{1em}
	\subfigure[A DT network with 1000 vertices]{
		\label{fig:3:b}       
		\includegraphics[width=0.46\textwidth]{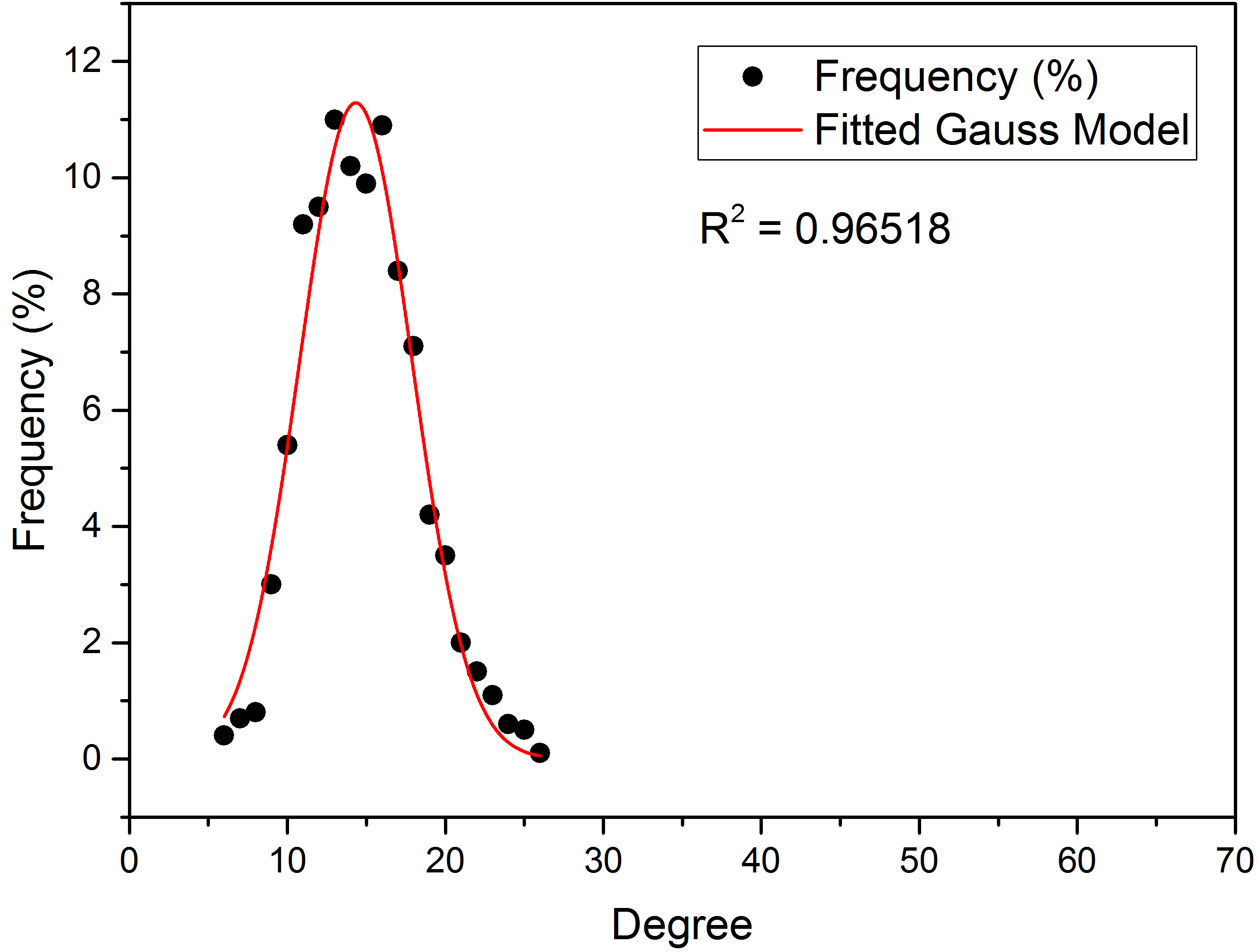}
	}
	\subfigure[A DT network with 10,000 vertices]{
		\label{fig:3:c}       
		\includegraphics[width=0.46\textwidth]{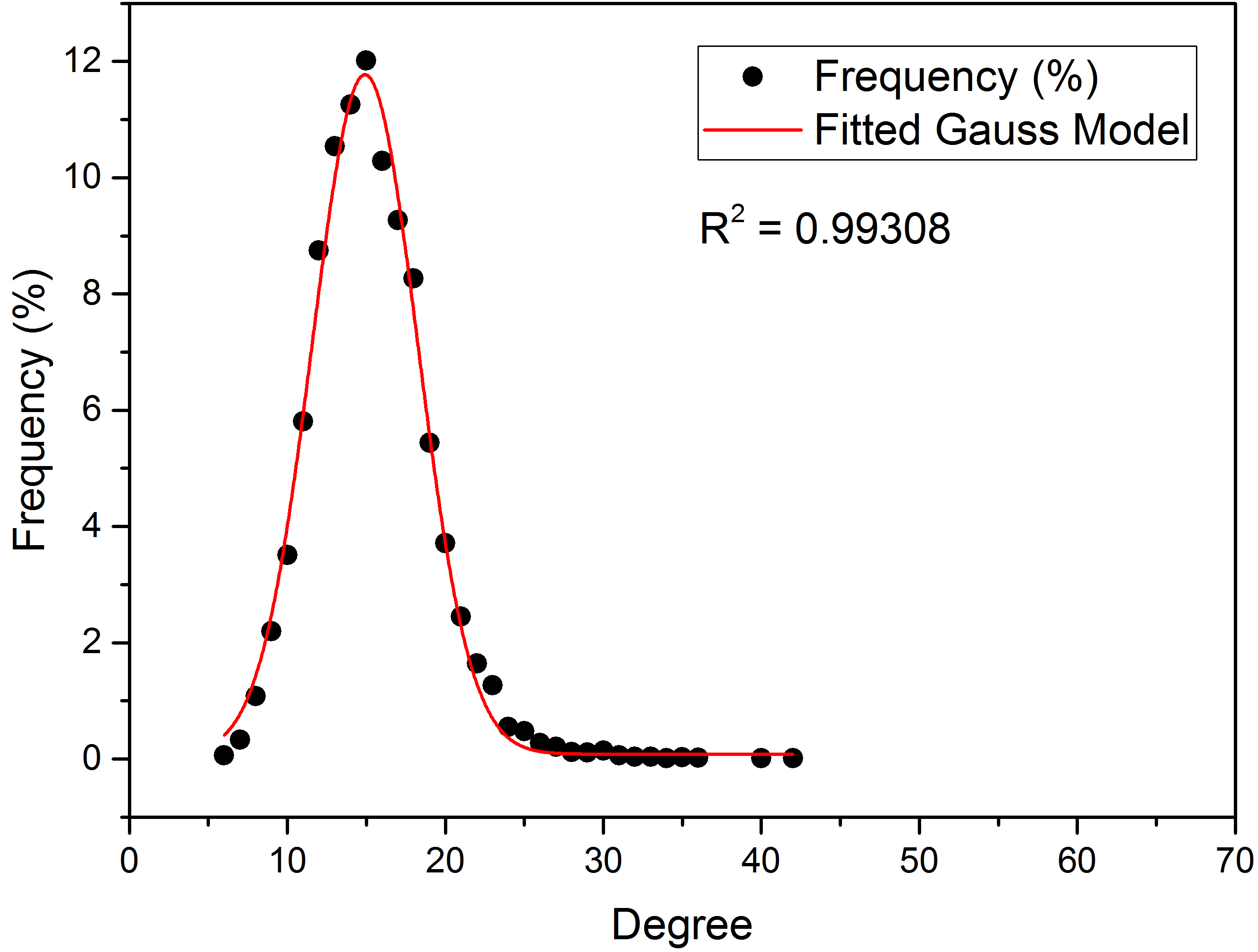}
	}
	\hspace{1em}
	\subfigure[A DT network with 100,000 vertices]{
		\label{fig:3:d}       
		\includegraphics[width=0.46\textwidth]{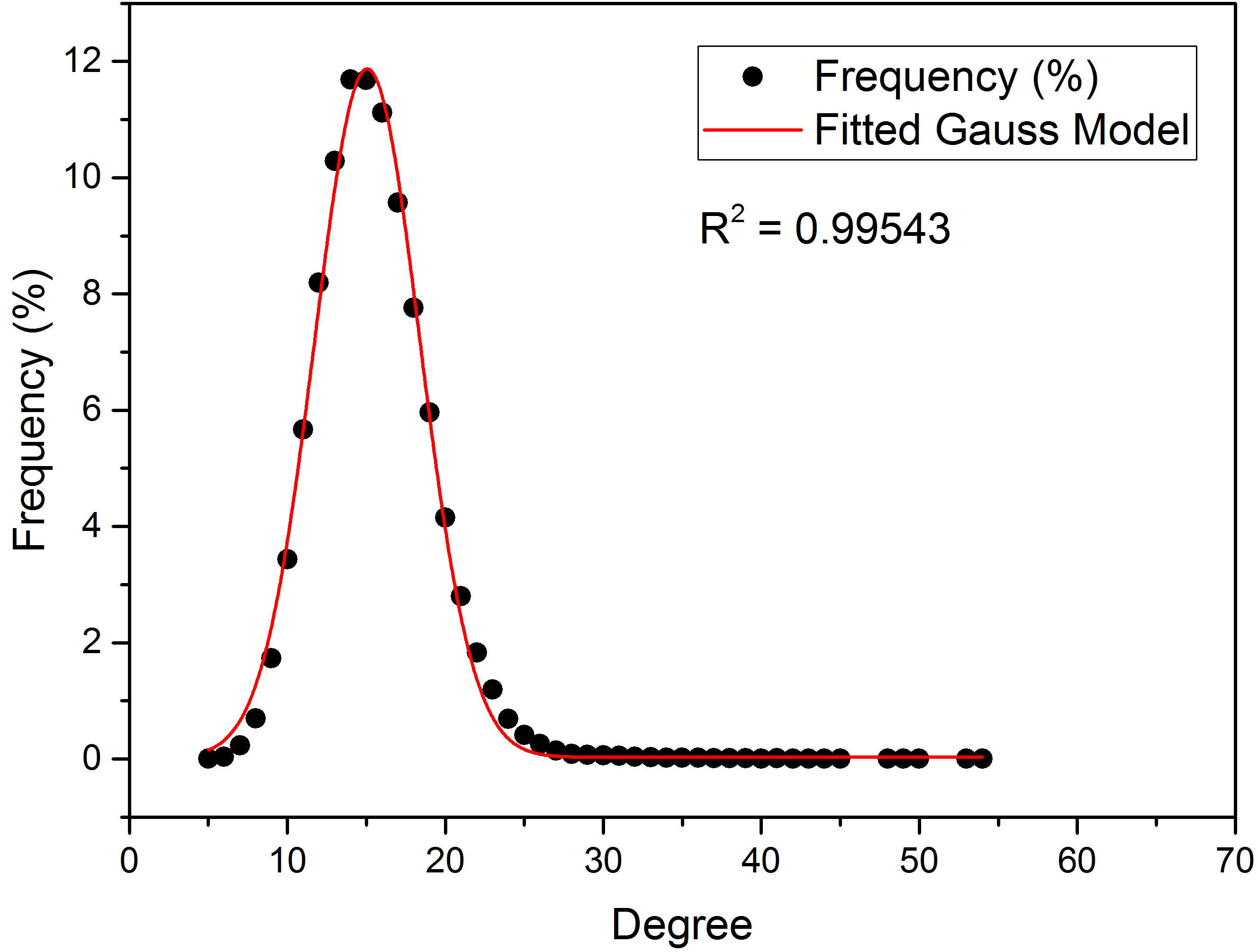}
	}
	\subfigure[A DT network with 1,000,000 vertices]{
		\label{fig:3:e}       
		\includegraphics[width=0.46\textwidth]{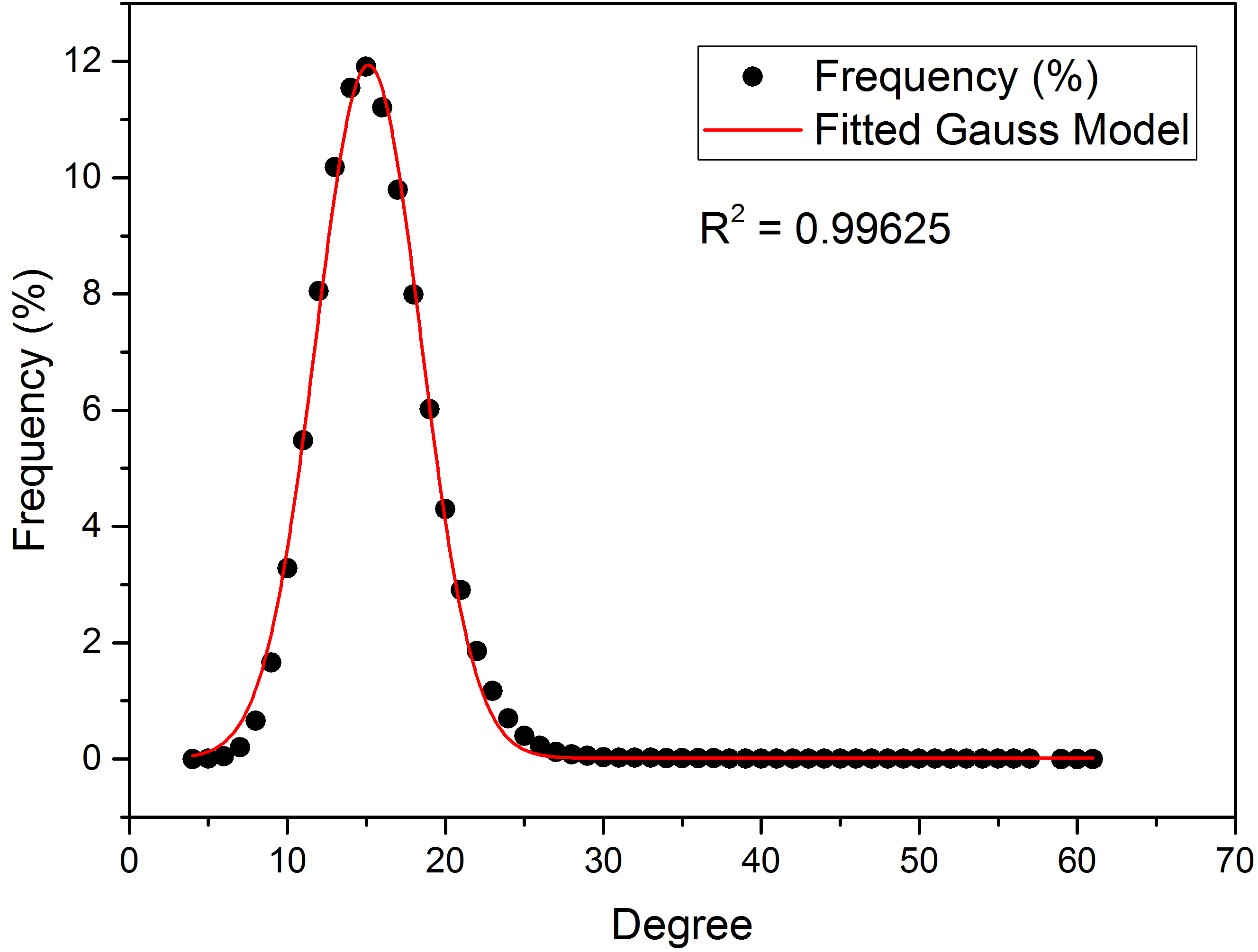}
	}
	\caption{Fitted Gaussian models for five different sizes of 
		three-dimensional DT networks}
	\label{fig:3}       
\end{figure}

\subsection{Degree Distribution of Three-dimensional DT Networks}

As the same as the investigation of the degree distribution of 
two-dimensional DT networks, for the three-dimensional ones, we also 
determine five different sizes of sets of scattered points. For each size, 
we repeatedly create the Delaunay tetrahedral meshes five times. We 
calculate the frequency of vertex degree, and then fit the frequency of 
vertex degree to the Gaussian function (Eq. (\ref{eq1})). Details of the fitted 
Gaussian Models are listed in Table \ref{tab2}. For each size, we also plot one of 
the fitted Gaussian Models (see Figure \ref{fig:3}). 

The results presented in Table \ref{tab2} and Figure \ref{fig:3} indicate that: 

\begin{enumerate}[label=(\arabic*), leftmargin=*]
	\item the degree distribution of three-dimensional DT networks also well follows the Gaussian Distribution except for the very small size of DTs (see Figure \ref{fig:3:a});
	\item with the increase of the number of vertices in DT networks, the frequency of vertex degree becomes more and more fitted to a Gaussian function. This is the same as that for the two-dimensional DT networks.
\end{enumerate}


\begin{table}[htpb] \small
	\caption{Details of the fitted Gaussian models for five different sizes of two-dimensional DT networks}
	\begin{center}
		\begin{tabular}{|p{35pt}|p{35pt}|p{35pt}|p{35pt}|p{35pt}|p{35pt}|p{35pt}|p{46pt}|}
			\hline
			\multirow{2}{2cm}{\textbf{Num. of Vertices}}& 
			\multicolumn{4}{|c|}{\textbf{Parameters of the Fitted Gauss Model}} & 
			\multicolumn{2}{|c|}{\textbf{Goodness of Fit}} & 
			\multirow{2}{*}{\textbf{Figure}} \\
			\cline{2-7} 
			& 
			\textbf{a}& 
			\textbf{b}& 
			\textbf{c}& 
			\textbf{y}$_{0}$ & 
			\textbf{R}$^{2}$ & 
			\textbf{Adj. R}$^{2}$ & 
			\\
			\hline
			\multirow{5}{*}{10$^{2}$}& 
			28.00545 & 
			5.41927 & 
			1.40459 & 
			0.38407 & 
			0.92611& 
			0.87069& 
			Figure 2(a) \\
			\cline{2-8} 
			& 
			29.68097 & 
			5.44527 & 
			1.27589 & 
			0.74051 & 
			0.93000& 
			0.87750& 
			\\
			\cline{2-8} 
			& 
			34.52490 & 
			5.98086 & 
			0.87684 & 
			4.03527 & 
			0.96943& 
			0.92357& 
			\\
			\cline{2-8} 
			& 
			34.08906 & 
			5.50212 & 
			0.98915 & 
			1.94301 & 
			0.97739& 
			0.96043& 
			\\
			\cline{2-8} 
			& 
			25.83382 & 
			5.41456 & 
			1.34654 & 
			1.75594 & 
			0.99149& 
			0.98510& 
			\\
			\hline
			\multirow{5}{*}{10$^{3}$}& 
			29.93515 & 
			5.81252 & 
			1.30081 & 
			0.31490 & 
			0.98151& 
			0.97041& 
			Figure 2(b) \\
			\cline{2-8} 
			& 
			29.93834 & 
			5.72799 & 
			1.20473 & 
			0.98548 & 
			0.98386& 
			0.97579& 
			\\
			\cline{2-8} 
			& 
			29.76963 & 
			5.76064 & 
			1.27671 & 
			0.51488 & 
			0.98669& 
			0.98004& 
			\\
			\cline{2-8} 
			& 
			29.16937 & 
			5.80171 & 
			1.35679 & 
			0.15905 & 
			0.98987& 
			0.98379& 
			\\
			\cline{2-8} 
			& 
			29.90300 & 
			5.84234 & 
			1.29982 & 
			0.29798 & 
			0.99680& 
			0.99519& 
			\\
			\hline
			\multirow{5}{*}{10$^{4}$}& 
			29.88082 & 
			5.81242 & 
			1.31279 & 
			0.16539 & 
			0.99024& 
			0.98699& 
			Figure 2(c) \\
			\cline{2-8} 
			& 
			29.93680 & 
			5.82429 & 
			1.30700 & 
			0.19757 & 
			0.99281& 
			0.99011& 
			\\
			\cline{2-8} 
			& 
			30.25405 & 
			5.82634 & 
			1.29838 & 
			0.11202 & 
			0.99342& 
			0.99145& 
			\\
			\cline{2-8} 
			& 
			30.38152 & 
			5.83662 & 
			1.28568 & 
			0.22451 & 
			0.99420& 
			0.99172& 
			\\
			\cline{2-8} 
			& 
			29.89607 & 
			5.84881 & 
			1.31856 & 
			0.12675 & 
			0.99494& 
			0.99326& 
			\\
			\hline
			\multirow{5}{*}{10$^{5}$}& 
			30.06407 & 
			5.83894 & 
			1.31350 & 
			0.09805 & 
			0.99287& 
			0.99093& 
			Figure 2(d) \\
			\cline{2-8} 
			& 
			30.16499 & 
			5.84439 & 
			1.30954 & 
			0.09481 & 
			0.99375& 
			0.99205& 
			\\
			\cline{2-8} 
			& 
			30.19954 & 
			5.84569 & 
			1.30756 & 
			0.09675 & 
			0.99415& 
			0.99255& 
			\\
			\cline{2-8} 
			& 
			30.07007 & 
			5.84758 & 
			1.31579 & 
			0.07981 & 
			0.99420& 
			0.99275& 
			\\
			\cline{2-8} 
			& 
			30.13382 & 
			5.84836 & 
			1.31393 & 
			0.07511 & 
			0.99428& 
			0.99285& 
			\\
			\hline
			\multirow{5}{*}{10$^{6}$}& 
			30.18968 & 
			5.84948 & 
			1.31402 & 
			0.05053 & 
			0.99423& 
			0.99315& 
			Figure 2(e) \\
			\cline{2-8} 
			& 
			30.08914 & 
			5.84862 & 
			1.31909 & 
			0.04873 & 
			0.99424& 
			0.99316& 
			\\
			\cline{2-8} 
			& 
			30.18773 & 
			5.85070 & 
			1.31355 & 
			0.05832 & 
			0.99427& 
			0.99305& 
			\\
			\cline{2-8} 
			& 
			30.21144 & 
			5.84841 & 
			1.31170 & 
			0.05835 & 
			0.99427& 
			0.99313& 
			\\
			\cline{2-8} 
			& 
			30.23149 & 
			5.84921 & 
			1.31112 & 
			0.05422 & 
			0.99432& 
			0.99326& 
			\\
			\hline
		\end{tabular}
		\label{tab1}
	\end{center}
\end{table}

\begin{table}[htpb] \small
	\caption{Details of the fitted Gaussian models for five different sizes of three-dimensional DT networks}
	\begin{center}
			\begin{tabular}{|p{35pt}|p{35pt}|p{35pt}|p{35pt}|p{35pt}|p{35pt}|p{35pt}|p{46pt}|}
				\hline
				\multirow{2}{2cm}{\textbf{Num. of Vertices}}& 
				\multicolumn{4}{|c|}{\textbf{Parameters of the Fitted Gauss Model}} & 
				\multicolumn{2}{|c|}{\textbf{Goodness of Fit}} & 
				\multirow{2}{*}{\textbf{Figure}} \\
			\cline{2-7} 
			& 
			\textbf{a}& 
			\textbf{b}& 
			\textbf{c}& 
			\textbf{y}$_{0}$ & 
			\textbf{R}$^{2}$ & 
			\textbf{Adj. R}$^{2}$ & 
			\\
			\hline
			\multirow{5}{*}{10$^{2}$}& 
			13.12919 & 
			12.86086 & 
			2.88146 & 
			0.40405 & 
			0.72202& 
			0.65253& 
			Figure 3(a) \\
			\cline{2-8} 
			& 
			14.08050 & 
			11.59738 & 
			1.88705 & 
			2.22933 & 
			0.79719& 
			0.74188& 
			\\
			\cline{2-8} 
			& 
			13.46764 & 
			12.53566 & 
			2.87983 & 
			0.52913 & 
			0.82998& 
			0.77331& 
			\\
			\cline{2-8} 
			& 
			13.12995 & 
			12.38275 & 
			2.85504 & 
			0.46392 & 
			0.83969& 
			0.79597& 
			\\
			\cline{2-8} 
			& 
			11.77248 & 
			11.92589 & 
			3.00907 & 
			1.35579 & 
			0.91722& 
			0.89239& 
			\\
			\hline
			\multirow{5}{*}{10$^{3}$}& 
			11.28722 & 
			14.33136 & 
			3.55809 & 
			0.00020 & 
			0.96518& 
			0.95904& 
			Figure 3(b) \\
			\cline{2-8} 
			& 
			11.64593 & 
			14.11783 & 
			3.05218 & 
			0.46269 & 
			0.96882& 
			0.96414& 
			\\
			\cline{2-8} 
			& 
			12.10603 & 
			14.39760 & 
			3.14854 & 
			0.19478 & 
			0.97574& 
			0.97210& 
			\\
			\cline{2-8} 
			& 
			11.34799 & 
			14.36068 & 
			3.42533 & 
			0.13816 & 
			0.97641& 
			0.97248& 
			\\
			\cline{2-8} 
			& 
			11.16427 & 
			14.30988 & 
			3.36066 & 
			0.28659 & 
			0.98464& 
			0.98234& 
			\\
			\hline
		\multirow{5}{*}{10$^{4}$}& 
			11.68965 & 
			14.90960 & 
			3.32630 & 
			0.08346 & 
			0.99308& 
			0.99236& 
			Figure 3(c) \\
			\cline{2-8} 
			& 
			11.56788 & 
			14.88169 & 
			3.39803 & 
			0.05132 & 
			0.99342& 
			0.99276& 
			\\
			\cline{2-8} 
			& 
			11.47522 & 
			14.86408 & 
			3.42240 & 
			0.04812 & 
			0.99346& 
			0.99287& 
			\\
			\cline{2-8} 
			& 
			11.69194 & 
			14.83496 & 
			3.31919 & 
			0.08442 & 
			0.99356& 
			0.99293& 
			\\
			\cline{2-8} 
			& 
			11.63035 & 
			14.79683 & 
			3.32669 & 
			0.09591 & 
			0.99378& 
			0.99316& 
			\\
			\hline
		\multirow{5}{*}{10$^{5}$}& 
			11.84302 & 
			15.05391 & 
			3.32420 & 
			0.03019 & 
			0.99543& 
			0.99510& 
			Figure 3(d) \\
			\cline{2-8} 
			& 
			11.83560 & 
			15.05495 & 
			3.32932 & 
			0.02664 & 
			0.99577& 
			0.99547& 
			\\
			\cline{2-8} 
			& 
			11.86178 & 
			15.06382 & 
			3.32584 & 
			0.02517 & 
			0.99583& 
			0.99554& 
			\\
			\cline{2-8} 
			& 
			11.85270 & 
			15.05600 & 
			3.32674 & 
			0.02680 & 
			0.99602& 
			0.99574& 
			\\
			\cline{2-8} 
			& 
			11.86734 & 
			15.05052 & 
			3.31587 & 
			0.03080 & 
			0.99603& 
			0.99574& 
			\\
			\hline
		\multirow{5}{*}{10$^{6}$}& 
			11.92144 & 
			15.13136 & 
			3.32073 & 
			0.01385 & 
			0.99625& 
			0.99604& 
			Figure 3(e) \\
			\cline{2-8} 
			& 
			11.93135 & 
			15.13290 & 
			3.31960 & 
			0.01300 & 
			0.99628& 
			0.99607& 
			\\
			\cline{2-8} 
			& 
			11.92786 & 
			15.13621 & 
			3.31965 & 
			0.01348 & 
			0.99638& 
			0.99617& 
			\\
			\cline{2-8} 
			& 
			11.92595 & 
			15.13733 & 
			3.31972 & 
			0.01397 & 
			0.99639& 
			0.99619& 
			\\
			\cline{2-8} 
			& 
			11.92224 & 
			15.13856 & 
			3.32184 & 
			0.01250 & 
			0.99647& 
			0.99628& 
			\\
			\hline
		\end{tabular}
		\label{tab2}
	\end{center}
\end{table}

\section{Conclusion}

In this paper, we first considered the two- and three-dimensional DTs as a 
type of complex networks, i.e., the DT network; and then we investigated the 
degree distribution of the DT networks. It has been found that the degree 
distribution of the DT networks well follows the Gaussian distribution in 
most cases. This is different from the Power-Law degree distributions in the 
Scale-Free networks and the Poisson degree distribution in the Small-World 
networks. 

In real-world applications, perhaps there are few complex systems can be 
exactly represented by the DT network. However, it is possible that in some 
cases scientists may try to use the DT network to approximately represent 
the quite complex biological network in the brain, and reveal the structural 
correlations of the biological networks \cite{11,12,13}. 

\section*{Acknowledgments}
This work was supported by the Natural Science 
Foundation of China (Grant Numbers 11602235 and 41772326).

%



\begin{thebibliography}{99}
	
\bibitem{1} Savelli, F. and J.J. Knierim, \textit{AI mimics brain codes for navigation.} Nature, 2018. \textbf{557}(7705): p. 313-314.

\bibitem{2}  LeCun, Y., Y. Bengio, and G. Hinton, \textit{Deep learning.} Nature, 2015. \textbf{521}: p. 436.

\bibitem{3}  Banino, A., et al., \textit{Vector-based navigation using grid-like representations in artificial agents.} Nature, 2018.

\bibitem{4}  Hafting, T., et al., \textit{Microstructure of a spatial map in the entorhinal cortex.} Nature, 2005. \textbf{436}: p. 801.

\bibitem{5} Doeller, C.F., C. Barry, and N. Burgess, \textit{Evidence for grid cells in a human memory network.} Nature, 2010. \textbf{463}: p. 
657.

\bibitem{6} Watts, D.J. and S.H. Strogatz, \textit{Collective dynamics of 'small-world' networks.} Nature, 1998. \textbf{393}(6684): p. 
440-442.

\bibitem{7} Barabasi, A.L. and R. Albert, \textit{Emergence of scaling in random networks.} Science, 1999. \textbf{286}(5439): p. 
509-512.

\bibitem{8} Delaunay, B., \textit{Sur la sphère vide.} Bulletin de l'Académie des Sciences de l'URSS, Classe des 
sciences mathématiques et naturelles, 1934. \textbf{6}: p. 793--800.

\bibitem{9} Shewchuk, J.R., \textit{Delaunay refinement algorithms for triangular mesh generation.} Computational Geometry-Theory and Applications, 2002. 
\textbf{22}(1-3): p. 21-74.

\bibitem{10} Si, H., \textit{TetGen, a Delaunay-Based Quality Tetrahedral Mesh Generator.} Acm Transactions on Mathematical Software, 2015. 
\textbf{41}(2).

\bibitem{11} Rubinov, M. and O. Sporns, \textit{Complex network measures of brain connectivity: Uses and interpretations.} NeuroImage, 2010. \textbf{52}(3): p. 
1059-1069.

\bibitem{12} Bullmore, E. and O. Sporns, \textit{Complex brain networks: graph theoretical analysis of structural and functional systems.} Nature Reviews Neuroscience, 2009. 
\textbf{10}: p. 186.

\bibitem{13} Avena-Koenigsberger, A., B. Misic, and O. Sporns, \textit{Communication dynamics in complex brain networks.} Nature Reviews 
Neuroscience, 2018. \textbf{19}(1): p. 17-33.
\end{thebibliography}

\end{document}